\documentclass[twocolumn,floatfix, showpacs]{revtex4}

\usepackage[final]{epsfig}
\usepackage{amsmath}
\begin{document}
 
\title{Jet-mass Dependence of the in-Medium Shower Modification}
 
\author{Thorsten Renk}
\email{thorsten.i.renk@jyu.fi}
\affiliation{Department of Physics, P.O. Box 35, FI-40014 University of Jyv\"askyl\"a, Finland}
\affiliation{Helsinki Institute of Physics, P.O. Box 64, FI-00014 University of Helsinki, Finland}


\begin{abstract}
While the modification of jets by a medium as created in ultrarelativistic heavy-ion collisions is often thought of as a phenomenon affecting a hard near on-shell parton and hence treated as partonic energy loss, a more realistic view is to think of the medium effect as a relatively small correction to the Quantum Chromodynamics (QCD) vacuum radiation pattern which reduces the initial high virtuality of hard partons and leads to the development of a shower. The uncertainty relation can be used to argue that a significant part of this shower is developed even before a medium can be formed. The initial virtuality of a shower-initiating parton is reflected in the final state in the measured invariant mass of a jet --- high mass jets are characterized by more branchings, higher multiplicity and a wider angular structure. Since most of this structure is determined before the medium affects the jet, selecting a particular jet mass range strongly biases the partonic configuration that enters the medium, and thus the medium is expected to modify high mass jets more strongly than low mass jets. In this work, this scenario is explored using the in-medium shower evolution code YaJEM and a possible strategy for a measurement is suggested.
\end{abstract}
 
\maketitle

\section{Introduction}

A hard QCD process typically results in highly virtual final state partons which subsequently shed virtuality in a characteristic radiation pattern accessible in perturbative QCD (pQCD) and evolve into parton showers which after hadronization become experimentally accessible as a jets, i.e. collimated sprays of hadrons. In a virtuality-ordered shower description such as e.g. the PYSHOW algorithm \cite{PYSHOW}, the radiation pattern is generated as a series of $1 \rightarrow2$ splittings $a\rightarrow b,c$ where the parton virtualities decrease $Q_b, Q_c \ll Q_a$. Energy-momentum balance then dictates that the lost parton virtuality leads to a transverse momentum difference between the daughter partons $b$ and $c$, however the invariant mass of the whole system is conserved. The branchings terminate as soon as parton virtuality reach a non-perturbative scale $O(1)$ GeV, at which point pQCD is no longer applicable. The virtuality $Q_i$ of the shower initiating parton thus governs both the average number of steps in the branching chain and the transverse momentum distribution of the jet fragments.

Experimentally jets are defined by applying a clustering algorithm to the final state hadrons, which typically involves (explicitly or implicitly) an angular distance $R$ of hadrons with respect to the jet axis and a minimum constituent transverse momentum $P_T$ cut as well as possible PID effects (the latter two reflecting the fact that not all hadron species reach the calorimeter or leave a signal). This means that the strict relation between jet invariant mass $M_{jet}$ and initial virtuality $Q_i$s broken --- hadrons at large angles $\phi > R$ with the jet axis carry the imprint of the original parton virtuality but are not counted as part of the jet and hence do not contribute to jet mass. This implies that any realistic calculation can not use $Q_i$ as a theoretical proxy for $M_{jet}$ but must take into account the clustering procedure. 

The uncertainty relation allows to estimate the timescale at which a splitting takes place as $E/Q^2$ with $E$ the intermediate parton energy scale. Since the initial virtuality is of order of the hard momentum scale $Q_i \sim O(p_T) \sim O(\text{100 GeV})$ for LHC kinematics, the timescale associated with the first splittings is parametrically $\sim O(0.01 \text{ fm}))$, i.e. far smaller than the medium formation time $\tau_0 \sim O(0.2-0.5 \text{ fm})$. What is potentially modified by the medium when the hard process takes place in a heavy-ion collision is thus the parton configuration as it has been created by the initial branchings by the time the medium is formed, and as discussed above, the nature of this configuration is correlated with $Q_i$ or $M_{jet}$.

The importance of going beyond an energy loss approximation for a leading on-shell parton and  computing this in-medium shower evolution has been pointed out already in \cite{Abhijit}. Intuitively one might expect that the medium modification of a shower depends on the multiplicity of showering partons by the time the medium is forming as each of the shower partons may interact with the medium, and thus the total momentum transfer between medium and shower partons scales with the number of available partons. As a result, one may expect that jets with large $M_{jet}$ show on average more substantial medium modifications than jets with small $M_{jet}$. In this work, we explore this idea using the in-medium shower evolution Monte Carlo (MC) code YaJEM \cite{YaJEM1,YaJEM2,YaJEM-D} in its latest version YaJEM-DE \cite{YaJEM-DE}, test if the effect is robust against the bias introduced by clustering and suggest possible experimental strategies to measure it.

\section{The model}

In order to illustrate the consequences of the jet mass effect most directly,  let us consider the simple case of a high $p_T$ quark with fixed energy undergoing shower evolution and fragmentation in the vacuum and in the presence of a medium. Experimentally, such a situation can be realized approximately in $\gamma$-h correlations where a trigger condition on the $\gamma$ restricts the momentum distribution of the recoiling parton (predominantly a quark) to a comparatively narrow region. We are interested in studying properties of the jets resulting from the QCD evolution of these quarks in vacuum and in medium.

In order to simulate the observable effects of $Q_i$, back-to-back quark pairs at given energy $E = 200$ GeV are generated as the final state of a hard process with the virtuality distribution of the quarks given by the initial splitting of the hard transient state into the final state of a back-to-back quark pair. One of the quarks is then selected and evolved with YaJEM down in virtuality to the hadron level. The resulting events are clustered using the anti-$k_T$ algorithm from the FastJet package \cite{FastJet} and binned according to $M_{J}$. We then study the longitudinal momentum distributions of hadrons inside the clustered jets in small and large $M_J$ bins and compute the ratio of the distributions which corresponds to $I_{AA}(P_T)$ in the experimental realization as $\gamma$-h correlation. In addition, we also study the transverse distribution of hadrons  $dN/d\phi(P_T)$ as a function of the angle with the jet axis $\phi$ in two different $P_T$ bins. The experimental observation would involve a $\gamma$-h coincidence using the jet axis that has been determined in the previous step (the need to do a correlation arises since the jet cone size $R$ otherwise restricts the angular region which can be probed). 

The core of the calculation, i.e. the evolution of parton showers in the medium is computed with the MC code YaJEM, which is based on the PYSHOW code \cite{PYSHOW} which in turn is part of PYTHIA \cite{PYTHIA}. It simulates the evolution from an  initial parton with virtuality $Q_i$  to a shower of partons at lower virtuality in the presence of a medium. In the absence of a medium, YaJEM by construction reproduces the results of PYSHOW. A detailed description of the model can be found in \cite{YaJEM1,YaJEM2,YaJEM-D}. Here we use the version YaJEM-DE \cite{YaJEM-DE} which is one of the best-tested theoretical models available for in-medium shower evolution and gives a fair account of a large number of high $P_T$ observables both at RHIC and LHC \cite{Constraining,A_J,RAA-LHC,A_J_edep}. 

In YaJEM-DE, the medium is characterized by two transport coefficients, $\hat{q}$ and $\hat{e}$. Here, $\hat{q}$ parametrizes the virtuality growth of a parton per unit pathlength and leads to medium-induced radiation whereas $\hat{e}$ describes the energy loss of propagating partons into non-perturbative medium modes. These transport coefficients are assumed to be related to the energy density $\epsilon$ of the hydrodynamical medium as

\begin{equation}
\label{E-qhat}
\hat{q}[\hat{e}](\zeta) = K[K_D] \cdot 2 \cdot [\epsilon(\zeta)]^{3/4} (\cosh \rho(\zeta) - \sinh \rho(\zeta) \cos\psi)
\end{equation}

with $\zeta$ the position along the path of the propagating parton, $\rho$ the medium transverse flow rapidity, $\psi$ the angle between parton direction and medium flow vector and $K,K_D$ two parameters regulating the strength of medium-induced radiation vs. direct energy loss into the medium. 
As in \cite{YaJEM-DE}, the free parameters $K, K_D$ are adjusted such that the energy loss into non-perturbative modes is a 10\% contribution to the total as constrained by a number of other observables \cite{Constraining}.

Given $(\hat{q}(\zeta),\hat{e}(\zeta))$ along the parton path, YaJEM computes the in-medium partonic shower evolution and hadronizes the result using the Lund string model \cite{Lund} so that the fragmentation remnants of the initial hard parton can be analyzed on the hadron level. However, a scaling law identified in \cite{YaJEM2} allows to simplify the computation by ordering shower according to their total line-integrated virtuality gain $\Delta Q^2_{tot} = \int d\zeta \hat{q}(\zeta)$, i.e. the precise functional form of $\hat{q}(\zeta)$ is significantly less important that the line-integrated value.

\section{Results}

\begin{figure}[!htb]
\begin{center}
\epsfig{file=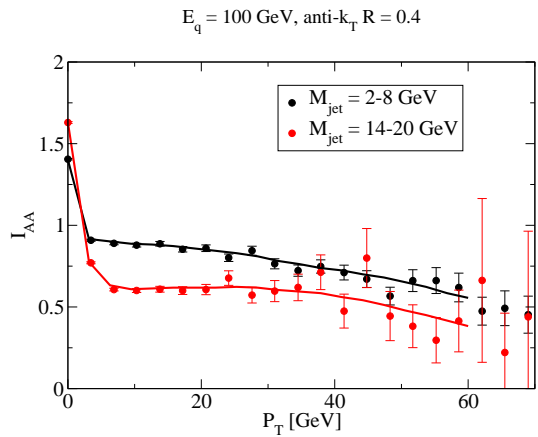, width=7cm}
\end{center}
\caption{\label{F-IAA}Ratio of medium-modified over vacuum fragmentation function for a 100 GeV quark for two different jet mass ranges (lines to guide the eye). Jets have been clustered with the anti-$k_T$ algorithm with $R=0.4$. }
\end{figure}

In Fig.~\ref{F-IAA}, the ratio of medium-modified over vacuum fragmentation functions (equivalently away side $I_{AA}$ in a $\gamma$-h correlation) is shown for a fixed $\Delta Q^2_{tot} = 10 $ GeV$^2$ which represents the medium modification for an average path in the medium as created at the LHC (see \cite{LHC-RAA} for details of the medium evolution model in which this is derived) in a low jet mass range $M_{jet} = 2-8$ GeV and a high jet mass range $M_{jet} = 14-20$ GeV.

(As a side remark, note that the use of the average $\Delta Q^2_{tot}$ is not justified in a detailed comparison with measured data which reflect the fluctuations around the average in a non-trivial way, but here it is a meaningful procedure to establish qualitative features and approximate magnitude of the effect. This is because unlike other observables which involve a convolution with the parton spectrum, such as the nuclear modification factor $R_{AA}$, the modification of the fragmentation function studied here is appoximately a linear function of $\Delta Q^2_{tot}$ and fluctuations can hence be expected to largely average out.)

It is clearly visible from the figure that, as expected, high mass jets which map to an increased shower activity before the medium modification subsequently also show stronger medium effects whereas low mass jets remain fairly robust against medium modification.

\begin{figure}[!htb]
\begin{center}
\epsfig{file=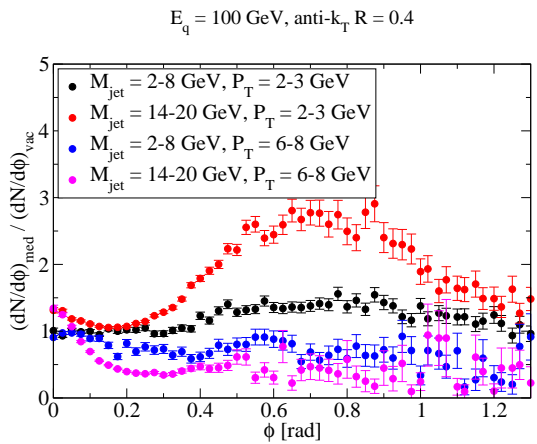, width=7cm}
\end{center}
\caption{\label{F-ang}Ratio of medium-modified over vacuum angular hadron distribution of a 100 GeV quark for two different jet mass and hadron $P_T$ ranges. Jets have been clustered with the anti-$k_T$ algorithm with $R=0.4$. }
\end{figure} 

A more differential look at ratios of the transverse medium and vacuum hadron distribution $dN/d\phi$ with respect to the jet axis in different $P_T$ bins as shown in Fig.~\ref{F-ang} shows the same pattern of suppression at high $P_T$ and enhancement at low $P_T$ as seen in Fig.~\ref{F-IAA}. However,
the effects are not unifrom in angle and the most dramatic differences between high and low jet mass and high and low $P_T$ window can be observed at $\phi > 0.4$ (i.e. outside the jet finding criterion used in this study so that a different experimental technique would be needed to study them). Again this is not unexpected as higher mass jets have more phase space available to populate large angles.

\section{Conclusions}

The simple study presented here establishes that there is reason to expect a non-trivial interplay between the vacuum radiation pattern as driven by the shower-initiating virtuality $Q_i$ and the subsequent medium-induced modification of the branching pattern: in a virtuality-ordered shower picture, showers get modified proportional to the number of virtual partons present by the time the medium is formed. An experimental analysis of observables differential in jet mass $M_{jet}$ is thus a non-trivial test for models of in-medium shower evolution and also in principle carries the imprint of the medium formation time.
The expected effects have been demonstrated to be robust against clustering (which tends to blur the relation between $Q_i$ and $M_{jet}$) and they appear to be large enough to be observed in practice through e.g. $\gamma$-jet correlations. Jet mass differential observables offer thus an exciting future opportunity for hard probes studies at RHIC and LHC.

\begin{acknowledgments}
 
I'm grateful to A.~Majumder who was the first to suggest this analysis to me. This work is supported by the Academy researcher program of the
Academy of Finland, Project No. 130472. 
 
\end{acknowledgments}

\end{document}